\documentclass[useAMS,usenatbib,usegraphicx]{mn2e}
\usepackage{amsmath}

\newif\ifAMStwofonts
\AMStwofontstrue

\newcommand{\simlt}{\lower.5ex\hbox{$\; \buildrel < \over \sim \;$}}
\newcommand{\simgt}{\lower.5ex\hbox{$\; \buildrel > \over \sim \;$}}

\title[A panchromatic analysis of starburst galaxy M82]
{A panchromatic analysis of starburst galaxy M82:\\
Probing the dust properties}

\author[Hutton et al.]
{Susan Hutton, Ignacio Ferreras\thanks{E-mail: i.ferreras@ucl.ac.uk}, Kinwah Wu,
Paul Kuin, \and Alice Breeveld, Vladimir Yershov, Mark Cropper, Mat Page\\
Mullard Space Science Laboratory, University College London, 
Holmbury St Mary, Dorking, Surrey RH5 6NT\\
}

\begin{document}
\date{MNRAS, in press.
Accepted 2014 January 24.  Received 2014 January 20; in original form 2013 July 5}
\pagerange{\pageref{firstpage}--\pageref{lastpage}} \pubyear{2014}
\maketitle
\label{firstpage}


\begin{abstract}
We combine NUV, optical and IR imaging of the nearby starburst galaxy
M82 to explore the properties of the dust both in the interstellar
medium of the galaxy and the dust entrained in the superwind. The
three NUV filters of {\sl Swift}/UVOT enable us to probe in detail the
properties of the extinction curve in the region around the 2175\,\AA\
bump. The NUV colour-colour diagram strongly rules out a ``bump-less''
Calzetti-type law, which can either reflect intrinsic changes in the
dust properties or in the star formation history compared to
starbursts well represented by such an attenuation law. We emphasize
that it is mainly in the NUV region where a standard Milky-Way-type
law is preferred over a Calzetti law.  The age and dust distribution
of the stellar populations is consistent with the scenario of an
encounter with M81 in the recent $\simlt 400$\,Myr. The radial
variation of NUV/optical/IR photometry in the galaxy region --
including the PAH-dominated emission at 8$\mu m$ -- confirms the
central location of the star formation. The radial gradients of the
NUV and optical colours in the superwind region support the
hypothesis that the emission in the wind cone is driven by scattering
from dust grains entrained in the ejecta.  The observed wavelength
dependence, $\propto \lambda^{-1.5}$, reveals either a grain size
distribution $n(a)\propto a^{-2.5}$, or a flatter distribution with a
maximum size cutoff, suggesting that only small grains are entrained
in the supernov\ae-driven wind.
\end{abstract}

\begin{keywords}
galaxies: individual (NGC\,3034) -- galaxies: evolution -- galaxies: starburst
-- galaxies: stellar content -- ISM: dust, extinction.
\end{keywords}

\section{Introduction}
\label{Sec:Intro}

Nearby star-forming galaxies provide the opportunity to probe our
knowledge of the mechanisms controlling star formation. Located at a
distance of 3.5\,Mpc \citep{Dalcanton:09}, and with a dynamical mass
of $\sim 10^{10}$M$_\odot$ \citep{Greco:12}, M82 (NGC\,3034) is the
closest starburst galaxy, and its edge-on orientation reveals a
prominent outflow powered by the accumulated energy injection from
supenov\ae\ into the interstellar medium \citep[see,
e.g.][]{MacLow:99}.
\citet{LS:1963} argued that M82 underwent an ``explosion'' leading to the
expulsion of gas along the minor axis.  The strong star formation
sustained by this galaxy is believed to have been triggered by a close
encounter with the more massive galaxy M81, located in the same
group. Dynamical modelling of the system suggests this encounter took
place between 300\,Myr \citep{Yun:93} and 1\,Gyr ago \citep{Sofue:98},
and there is a prominent distribution of gas and dust in the
intergalactic region between them \citep{Yun:93,Roussel:10}.

Multi-band photometry reveals a young population overall with a
luminosity-weighted age between 100 and 450\,Myr, and a central
(inside 500\,pc) region dominated by younger ($\sim$10\,Myr)
stars \citep{Mayya:06,RodMer:11}, possibly formed over a number of
episodic bursts lasting a few million years
each \citep{FS:03}. Detailed observations of star clusters in the
central region and throughout the disc support this
view \citep{Smith:06,Konst:09}. Furthermore, recent hydrodynamical
simulations suggest that many massive stellar clusters must have
formed over the past 10\,Myr in order to explain the multi-phase
properties of the wind \citep{Melioli:13}. Further out in the disc,
the central starburst may have also caused a decrease of the star
formation rate, as revealed by the low number of red
supergiants \citep{Davidge:08}.

The UV/optical/IR emission in the wind region extends out to
$\sim$10\,kpc \citep{Lehnert:99,Devine:99}.  The
polarization of the H$\alpha$
emission \citep[e.g.][]{VS:72,Yoshida:11} is explained by scattering
from dust particles entrained in the superwind \citep{SandBal:71}.
{\sl GALEX} UV imaging of the wind region confirms that neither
shock-heated nor photoionised gas can be the dominant mechanisms of
emission \citep{Hoopes:05}. The positive detection of Polycylic
Aromatic Hydrocarbons (PAH) in the infrared with {\sl
Spitzer} \citep{Engelbracht:06} and AKARI \citep{Kaneda:10}, and an
even more extended distribution of cool dust seen by {\sl
Herschel} \citep{Roussel:10}, give further support to this scenario,
although additional processes in the superwind have to be considered
when dealing with the emission
lines \citep{McKeith:95}. \citet{Ohyama:2002} split the wind region
into a diffuse component -- responsible for the scattered light -- and
a filamentary component, explained by shocks around the hot gas. In
addition to a supernov\ae-driven mechanism, \cite{Roussel:10} suggest
that over two-thirds of the dust in the intergalactic region could
have been removed from the M81/M82 system via tidal interactions.

In this paper, we take advantage of the increased spectral resolution
around the 2175\AA\ bump, provided by the {\sl Swift}/UVOT passbands,
in order to probe the dust properties in M82. In addition to exploring
the stellar populations in the galaxy, we extend the analysis to the
wind region by analysing the wavelength dependence of the light
scattered away from the galactic disc.  The structure of the paper is
as follows: the NUV-to-IR data used for the analysis are presented
in \S\ref{Sec:Data}, followed by the photometric analysis with
population synthesis and dust attenuation models in \S\ref{Sec:glx}.
In \S\ref{Sec:DustModel} we present a simple model to constrain the
properties of the dust entrained in the supernov\ae-driven
wind. Finally, our conclusions are given in \S\ref{Sec:Conc}.

 \begin{table}
 \caption{Log of {\sl Swift}/UVOT observations of NGC\,3034 used in this paper.}
 \label{tab:uvot}
 \begin{tabular}{ccrrr}
 \hline
 & & \multicolumn{3}{c}{Exposure Time (s)}\\
 ObsID & Date & UVW2 & UVM2 & UVW1\\
 & & 2033\AA & 2229\AA & 2591\AA\\
 \hline
 00031201001 & 2008/05/01 & 1660.68  &  1660.67  &  1469.57\\
 00031201002 & 2009/04/25 &   ---    &  4636.08  &   ---   \\
 00032503003 & 2012/07/06 &   ---    &   973.11  &   ---   \\
 00032503004 & 2012/07/13 &  954.40  &   ---     &   ---   \\
 00032503006 & 2012/07/27 &   ---    &   ---     &  1042.70\\
 00035482001 & 2007/01/26 & 1518.53  &   999.43  &   758.71\\
 00035482002 & 2009/10/18 & 1442.39  &  1038.59  &   719.94\\
 00091489001 & 2012/04/05 &  413.15  &   413.15  &   428.54\\
 00091489002 & 2012/04/06 & 1572.38  &  1572.53  &  1260.46\\
 00091489003 & 2012/04/08 & 1256.48  &  1256.47  &  1424.05\\
 00091489004 & 2012/04/10 &  289.15  &   289.15  &   263.69\\
 00091489005 & 2012/04/14 &  111.98  &   111.98  &   130.21\\
 00091489006 & 2012/04/15 &  875.34  &   875.41  &  1038.83\\
 00091489007 & 2012/04/17 &  118.87  &   118.86  &   159.45\\
 \hline
            & TOTAL      & 10213.35  & 13945.43  &  8696.15\\
 \hline
 \end{tabular}
 \end{table}

\begin{figure}
\includegraphics[width=8cm]{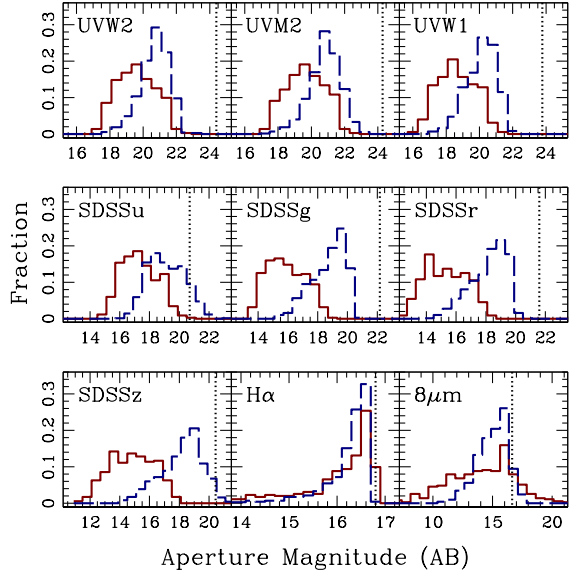}
\caption{Distribution of aperture magnitudes in the
galaxy (solid red histograms) and the wind region (blue dashed).  In
each panel, we also include the limiting magnitude derived from the
distribution of fluxes measured in 1000 random apertures in the
background (vertical dotted line). Note the histograms of the
GALEX/FUV and SDSS-i photometry are not included in this figure to
avoid overcrowding, see Tab.~\ref{tab:filters} for information about
these bands.}
\label{fig:MagLim}
\end{figure}

\begin{figure*}
\includegraphics[height=7cm]{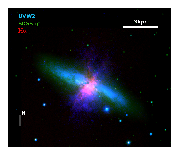}
\includegraphics[height=7cm]{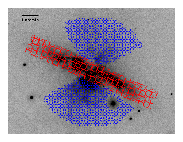}
\caption{{\sl Left:} RGB colour composite of M82 using UVOT/UVW2 (blue),
SDSS/g (green) and narrow band H$\alpha$ (red). Notice the 
wide extent of the NUV light in the wind region.
{\sl Right:} Apertures used for the analysis of the galaxy (red) and
the wind region (blue). Each aperture has a radius of 5\,arcsec. The greyscale
background image is that from the UVW2 filter.}
\label{fig:col}
\end{figure*}

\section{Data}
\label{Sec:Data}

The near ultraviolet science-grade images (NUV) of M82 were retrieved
from the \emph{Swift} archive at HEASARC\footnote{\tt
http://heasarc.gsfc.nasa.gov/docs/swift/}.  The Ultraviolet/Optical
Telescope \citep[UVOT,][]{Roming:05} is one of three telescopes
onboard the \emph{Swift} spacecraft \citep{Gehrels:04}. For our
purposes of targeting the dust properties of M82, UVOT has an optimal
set of three broadband filters in the NUV, straddling the 2175\AA\
bump \citep{Poole:08}. The camera covers a field of view of $17\times
17$\,arcmin$^2$, hence a single pointing covers the M82 and superwind
region, mapping a $18\times 18$\,kpc$^2$ area at the distance of
M82. The resolution of the telescope is between 2.4 and 2.9\,arcsec
(FWHM) in the NUV filters \citep{Breeveld:10}. The observational data
for this paper were taken between 2008 and 2012 (see
Tab.~\ref{tab:uvot} for details). The total exposure time amounts to
10.21, 13.95 and 8.70\,ks in the UVW2, UVM2 and UVW1 passbands,
respectively. All images were aspect corrected. In addition, the
standard corrections for large scale sensitivity \citep{Breeveld:11}
and the slow drift in zero point \citep{Breeveld:10b} were applied.
The individual exposures were registered and co-added using a standard
drizzle algorithm \citep{driz}, with a 0.5\,arcsec pixel size. 
Each ObsID set comprises several frames, so in total we have
around 50 frames in each band. Drizzling enables us to improve the S/N
ratio and the spatial resolution of the co-added image. Throughout
this paper we compute magnitudes within the standard 5\,arcsec
(radius) aperture.  We note this aperture is accurately
calibrated \citep{Poole:08,Breeveld:10} and minimizes the contribution
from the tails of the PSF. Table~\ref{tab:filters} summarises the
details of the passbands used in this paper, along with the
3\,$\sigma$ limiting aperture magnitudes. These magnitudes were
obtained by taking 1000 measurements within the same-sized circular
apertures, in blank outer regions of the images, away from the galaxy
or other contaminating sources.  The distribution of the aperture
magnitudes is shown in Fig.~\ref{fig:MagLim} both for the galaxy and
the wind footprint. Notice that the UVOT data are deep, so that the
photometry in the galaxy and wind regions is at least $2$\,mag
brighter than the 3\,$\sigma$ limit. Hence, the statistical
uncertainty of the photometric measurements remains below the 5\%
level. The combined effect of zero-point
uncertainties \citep[3\%;][]{Breeveld:11} and correction for Galactic
foreground dust (see below) results in a total uncertainty of
0.1\,mag. Regarding the Milky Way foreground correction, we take the
revised value of reddening from the NED database towards M82 
(E$_{\rm B-V}$=0.140\,Gyr), based on the
recalibration \citep{Schlafly:11} of the \citet{Schlegel:98} dust
maps. Along with the tabulated colour excess, we ran a large set of
synthetic population models from \citet{BC03} to compute the effect of
the stellar population properties on the correction for the different
passbands.  We use a wide range of ages and metallicities, applying
the standard R$_v$=3.1 Milky Way attenuation law \citep{Fitz:99}.  The
de-reddening correction (quoted in Tab.~\ref{tab:filters}) is defined
as the median value of the distribution for this range of models.  We
derive a systematic uncertainty from foreground dereddening of 6\% in
the UVOT UV filters (in contrast with the SDSS data, see below, where
the correction introduces an error $<$1\%).

\begin{table}
\caption{Properties of the passbands and limiting magnitudes}
\label{tab:filters}
\begin{tabular}{lcccc}
\hline
Filter  & $\langle\lambda\rangle$ & FWHM & Apert. Limit & MW Corr$^1$\\
        &    \AA   &  \AA   & 3\,$\sigma$(AB) & mag\\
\hline
\multicolumn{5}{c}{UVOT}\\
UVW2      &  2033 &   657 & 24.39 & 1.07\\
UVM2      &  2229 &   498 & 24.28 & 1.17\\
UVW1      &  2591 &   693 & 23.77 & 0.88\\
\multicolumn{5}{c}{GALEX}\\
FUV       &  1539 &   230 & 23.00 & 1.26\\
NUV       &  2316 &   793 & 23.96 & 1.27\\
\multicolumn{5}{c}{SDSS}\\
SDSS$u$   &  3551 &   599 & 20.70 & 0.67\\
SDSS$g$   &  4686 &  1379 & 22.17 & 0.52\\
SDSS$r$   &  6165 &  1382 & 21.59 & 0.36\\
SDSS$i$   &  7481 &  1535 & 20.62 & 0.27\\
SDSS$z$   &  8931 &  1370 & 20.44 & 0.20\\
\multicolumn{5}{c}{Other}\\
H$\alpha$ &  6573 &    67 & 16.80 & 0.36\\
IRAC/ch4  & 79274 & 28427 & 16.65 & 0.01\\
\hline
\end{tabular}
$^1$ Foreground Galactic attenuation correction towards M82
(see \S\ref{Sec:Data} for details).
\end{table}

In addition to the NUV data, we retrieved optical images in the $u$,
$g$, $r$, $i$, and $z$ bands from the DR8 version of the Sloan Digital
Sky Survey \citep{sdss:dr8}. In order to bring the spatial resolution
of the SDSS images (PSF FWHM $\simlt 1.5$\,arcsec) in line with the
UVOT data, we extracted an isolated stellar image from the UVOT UVW2
frame, and used a rescaled version of this PSF frame as a convolution
kernel for the SDSS images. The scale factor is chosen so that the PSF
of the convolved images matches the one corresponding to the NUV
data. Finally, the SDSS convolved frames were rescaled and registered
to match the UVOT images. By comparing photometric measurements before
and after this process, we estimate an additional 5-10\% error in the
optical fluxes. The SDSS data are deep enough for our purposes, except
for the wind region in the shallower $u$ band frame (see
Fig.~\ref{fig:MagLim}). Nevertheless, even in this case, only 11\% of
the measurements fall below this limit, all of them at a projected
galactocentric distance $R>2.6$\,kpc.

\begin{figure}
\includegraphics[width=8.5cm]{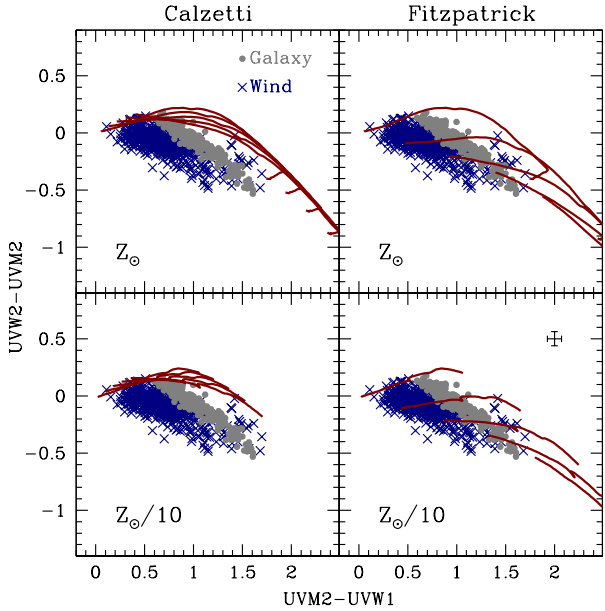}
\caption{
NUV colour-colour plots for the galaxy (grey filled circles) and wind
(blue crosses) regions, within a projected galactocentric distance of
3\,kpc. This figure compares the model predictions for two different
extinction laws: Calzetti (2001; {\sl left}) and Fitzpatrick (1999;
{\sl right}). Each line represents an age sequence from 0.1 to 10\,Gyr
for a simple stellar population from \citet{BC03} at solar metallicity
({\sl top}) and $Z_\odot/10$ ({\sl bottom}). The lines in each panel
corresponds to different values of the colour excess arising from the
extinction, from top to bottom, E$_{\rm B-V}=\{0,0.25, 0.50, 0.75,
1\}$\,mag. A characteristic $1\sigma$ error bar is shown in the
bottom-right panel.}
\label{fig:NUVccd}
\end{figure}

We retrieved and processed the H$\alpha$-subtracted image of M82 from
the fifth data release of the Spitzer Infrared Nearby Galaxies
Survey \citep[SINGS,][]{sings}. The image, taken at the 2.1\,m KPNO
telescope, was processed the same way as the SDSS frames, to match the
pixel size and resolution of the UVOT images. In addition, we made use
of the Spitzer/IRAC 8$\mu$m image of M82, also retrieved from the
SINGS data release.  Given the resolution of this image, we decided
only to register and re-scale the 8$\mu$m data to the reference frame.
Finally, the GALEX images were retrieved from the Atlas of Nearby
Galaxies \citep{GdP:07}. In this paper we use only the FUV data,
because the NUV photometry is comparable to UVOT/UVM2 \citep[see,
e.g.,][]{SUSS}.

Fig.~\ref{fig:col} ({\sl left}) shows a colour composite of our data,
with the blue, green, and red channels of the RGB image corresponding
to UVW2, SDSS $g$ and H$\alpha$, respectively. Note the significant
colour difference between the bulk of the galaxy -- mostly blue-green,
implying it is dominated by intermediate stellar ages -- and the central
region, showing intense H$\alpha$ and UV emission from young, massive
stars. In the outer regions, perpendicular to the plane of the disc,
the dominant emission is in the UV, reflecting the contribution from
dust scattering (see \S\ref{Sec:DustModel}).

\begin{figure}
\includegraphics[width=8.7cm]{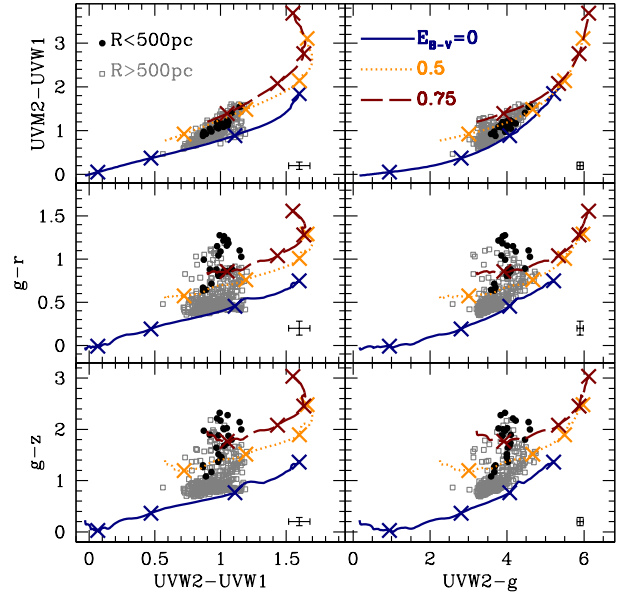}
\caption{NUV and optical colour-colour diagrams of the photometric data
in the galaxy region. The sample is split with respect to the
projected galactocentric distance (R), as labelled. For reference, an
age sequence at solar metallicity is shown, from the models
of \citet{BC03} assuming a dustless population (blue solid lines) and
two dusty cases (orange dotted lines for E$_{\rm B-V}$=0.5; and red
dashed lines for E$_{\rm B-V}$=0.75), following a \citet{Fitz:99}
extinction law.  The crosses mark, from left to right, stellar ages of
0.1, 0.5, 1 and 5\,Gyr.  A typical error bar is shown in each panel.}
\label{fig:CCD}
\end{figure}

From these images we derived a photometric catalogue, separately for the
galaxy and the superwind region, by selecting a number of 5\,arcsec
(radius) apertures, as shown on the right hand panel of
Fig.~\ref{fig:col}. We took care in removing apertures in the areas
contaminated by foreground/background sources. This is especially
important in the wind region, where bright sources would introduce a
number of outliers in the study of the photometry. The final data set
uses 351 (695) apertures in the galaxy (wind) footprint.

\begin{figure*}
\includegraphics[width=8.7cm]{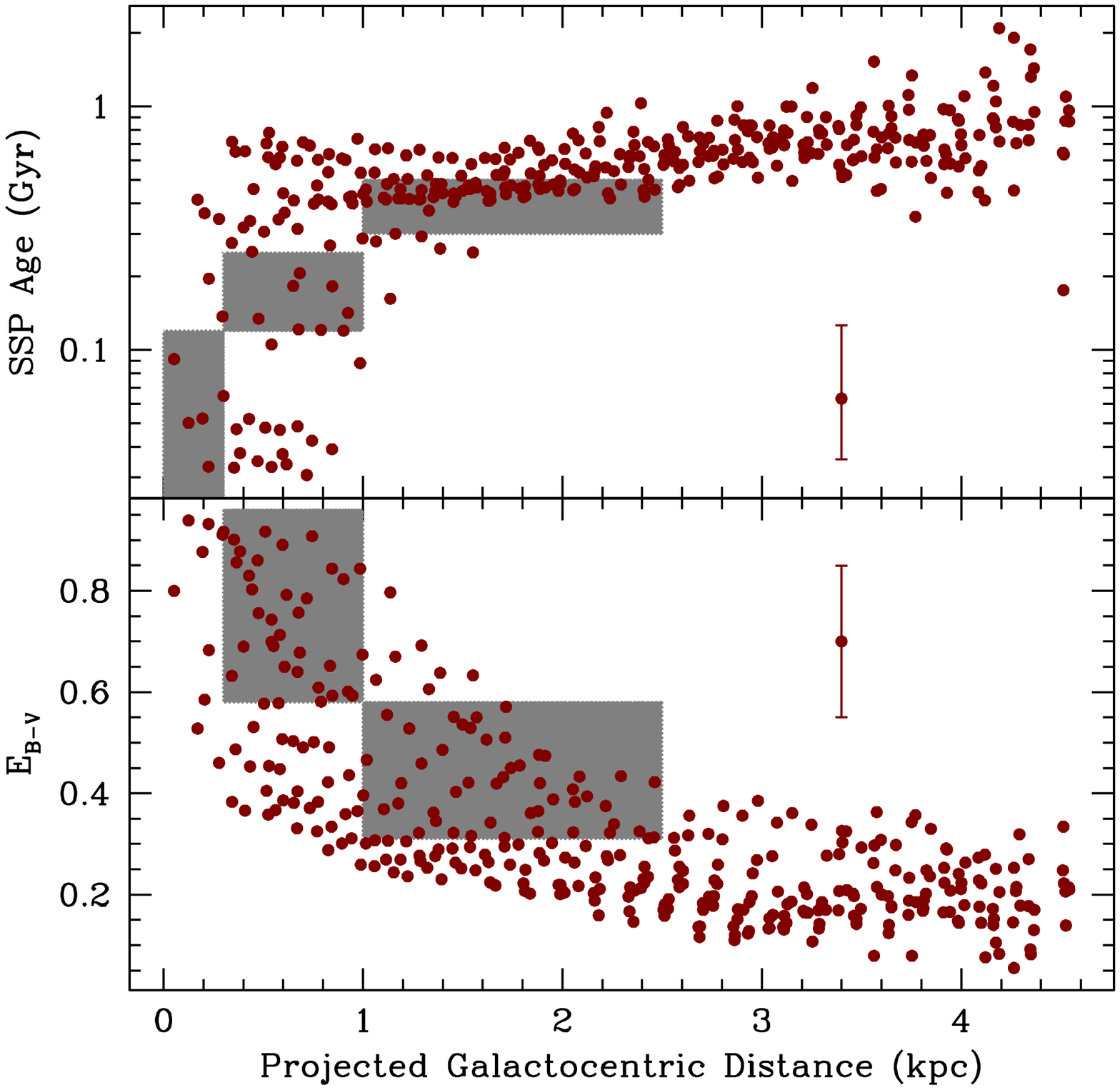}
\includegraphics[width=8.7cm]{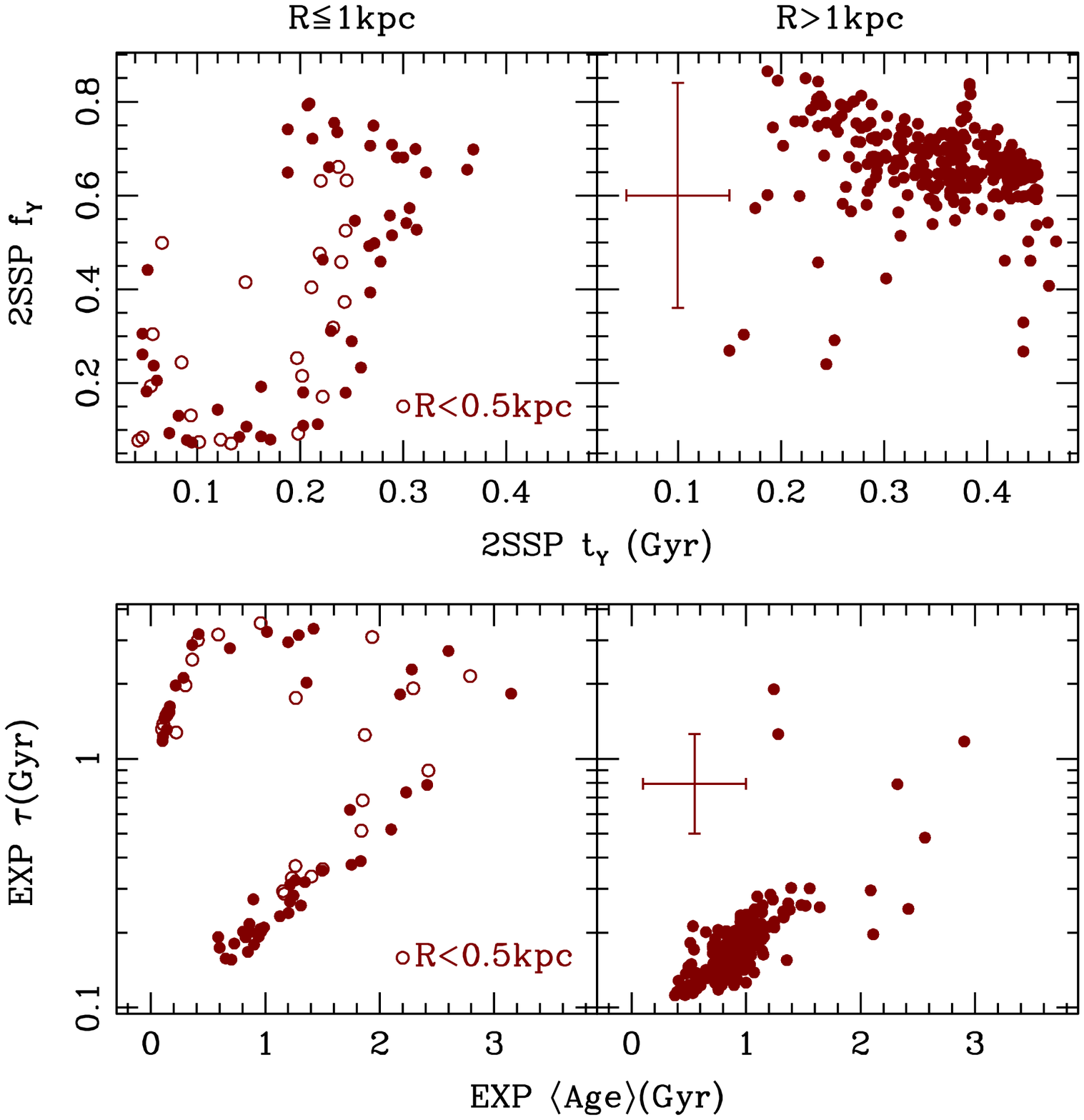}

\caption{
{\sl Left:} Best-fit SSP equivalent ages ({\sl top}) and colour excess
({\sl bottom}) of the data in the galaxy region, obtained by a
comparison with a set of 32,768 simple stellar populations from the
models of \citet{BC03} (see text for details). A characteristic
1\,$\sigma$ error bar is shown in each case. The grey regions
correspond to the range of ages and colour excess
from \citet{RodMer:11}.
{\sl Right:} Best fits for composite models: the top panels show
the values of the young component in age (t$_{\rm Y}$) and
stellar mass fraction (f$_{\rm Y}$) for the 2SSP models; whereas
the bottom panels correspond to the average age and timescale
for the exponentially decaying SFHs. The data are split with respect
to galactocentric distance, as labelled (top), and typical $1\,\sigma$ error
bars are included in the rightmost panels.}
\label{fig:ages}
\end{figure*}

\section{Photometry of the galaxy region}
\label{Sec:glx}

Given the passband response curves of the UVOT NUV filters
\citep[see, e.g.,][]{Breeveld:11}, a colour-colour diagram using these filters
is a powerful discriminant of the dust extinction spectral properties of
nearby galaxies, especially in the region around the
2175\,\AA\ bump. Although the origin of the bump is not clear, it
matches a resonance in transitions involving C-ring structures such as
graphite, or PAH compounds \citep[see, e.g.,][]{Duley:98}. Even though
this bump is strong in sightlines probing the ISM of the Milky Way
galaxy \citep{FM:86}, it seems to be absent in starburst galaxies
\citep{Calz:01}. The lack of a strong bump could be indicative
of changes in the dust properties \citep{GCW:97}. However, an
age-dependent attenuation law from an otherwise identical dust
component can also give rise to different bump
strengths \citep{Panuzzo:07}.

\subsection{NUV photometry and dust extinction}

Fig.~\ref{fig:NUVccd} shows an NUV colour-colour diagram with the
three UVOT filters. All four panels show the same photometric data,
and each panel overlays a different set of models. Each red line
tracks an age sequence corresponding to a simple stellar population
from the models of \citet{BC03}, at either solar metallicity
($Z_\odot$, {\sl top}) or $Z_\odot/10$ ({\sl bottom}). The lines span a
wide range of ages, from 0.1\,Gyr to 10\,Gyr (all models run with
age increasing from left to right).  Within each panel, the different
lines probe a range of reddening values, from a colour excess of
E$_{\rm B-V}$=0 (in the top of each panel) to 1\,mag, in steps of
$0.25$\,mag. Since these models only explore the evolution of a
synthetic stellar population, these tracks should be compared with the
photometry in the galaxy region (grey points). However, for reference,
we also show the photometric data in the wind region (blue crosses).
Finally, the models are also divided with respect to the extinction
law, using a \citet{Calz:01} prescription in the left-hand panels, and
a Milky-Way extinction law \citep{Fitz:99} on the right-hand panels.

Notice that although the models span a wide range of age, metallicity
and colour excess, the Calzetti extinction law ({\sl left}) cannot
account for the observations. A \citet{Fitz:99} law, however, covers
all the datapoints, suggesting that the extinction law of a starburst
galaxy such as M82 shows a prominent NUV bump. The UVW2$-$UVM2 and
UVM2$-$UVW1 colours straddle the 2175\,\AA\ bump, so this diagram is
especially constraining with respect to the presence of the bump.  We
show below that using models with composite stellar
populations still yields the same rejection of the Calzetti extinction
law in M82. Notice the increasingly blue value of the UVW2$-$UVM2
colour with increasing age. This is caused by the presence of a red
leak that introduces flux in the UVW2 filter for older
populations. We emphasize that the models take
into account the red leak.

\begin{figure*}
\raisebox{-0.5\height}{\includegraphics[width=8cm]{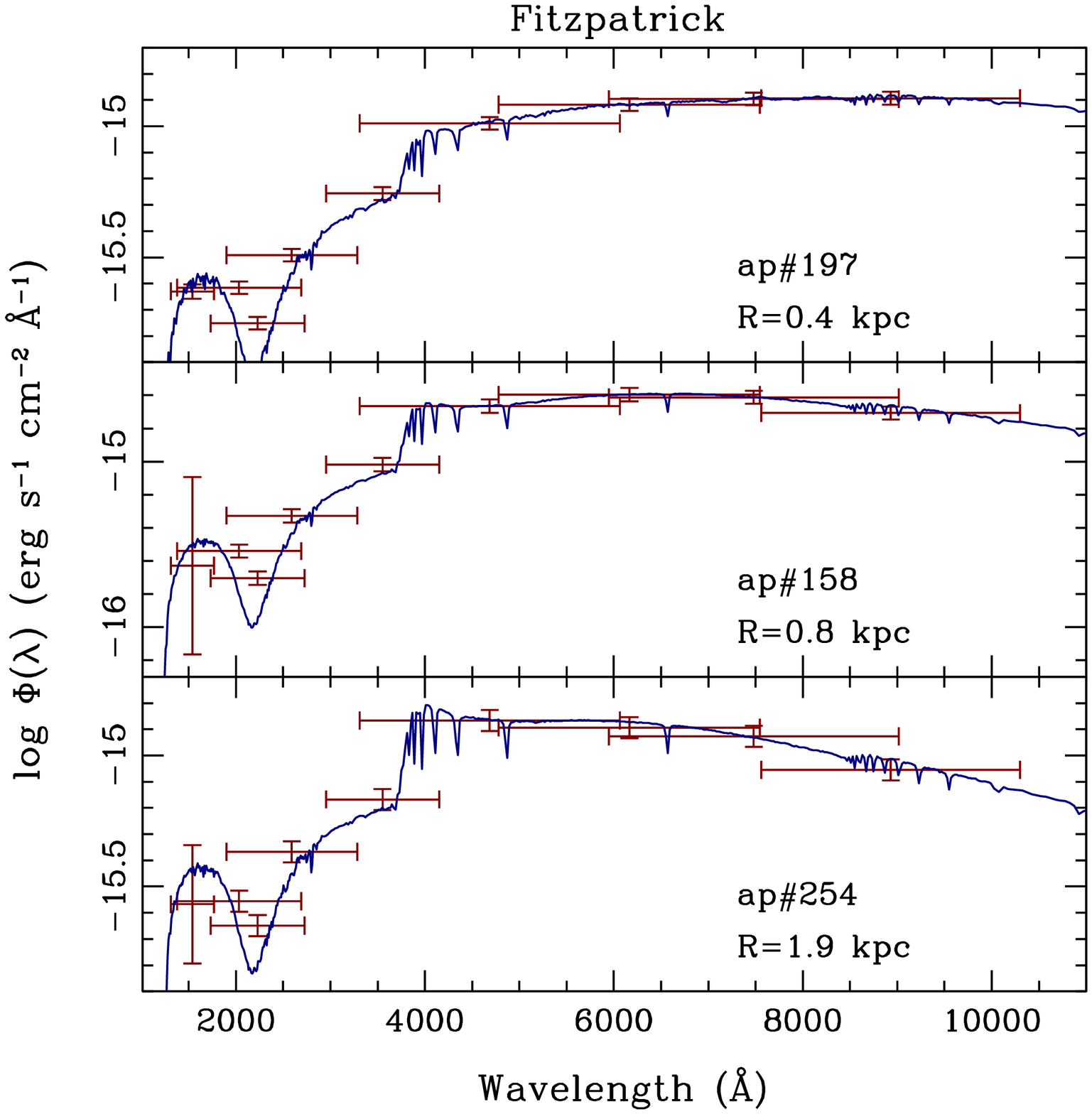}}
\raisebox{-0.5\height}{\includegraphics[width=8cm]{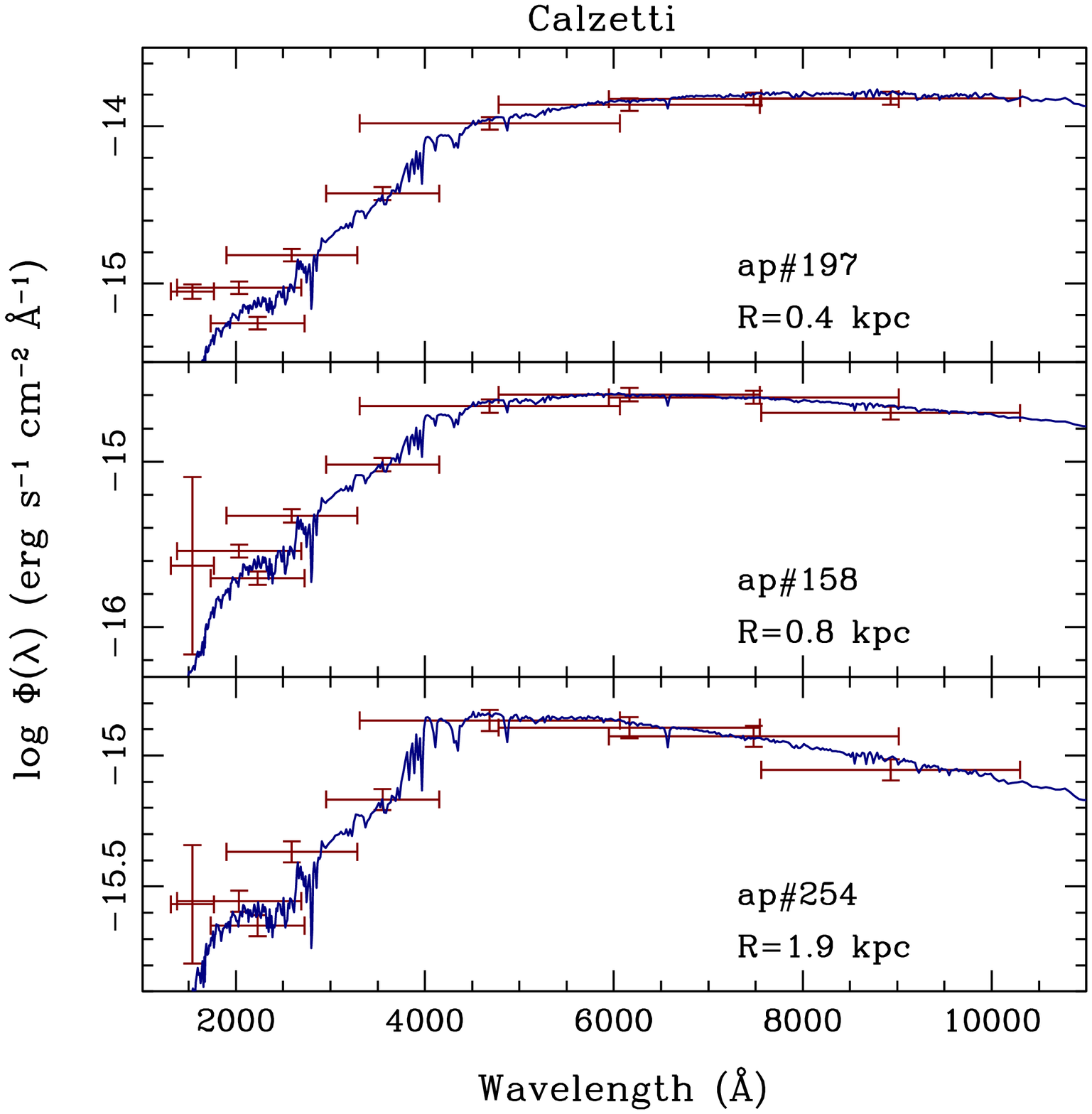}}

\caption{Comparison between observed NUV/Optical photometry (red error bars) 
and the best-fit models (blue) for three aperture measurements at 
different galactocentric distances, as labelled. The best fits are shown
assuming either a Fitpatrick ({\sl left}) or a Calzetti ({\sl right}) dust
extinction law. Note the significant difference between these two
in the NUV region, where the bump creates a dip in the Fitzpatrick case.}
\label{fig:SEDfit}
\end{figure*}

Even though the models cannot be compared with the photometric data in
the wind (blue crosses), this figure illustrates the significantly
bluer colours in this region.  The photometry in the wind region
reveals the presence of dust entrained in the gas ejected from a
supernov\ae -driven wind \citep{Hoopes:05}. This dust scatters light
from the central starburst. A detailed photometric analysis allows us
to constrain the properties of the dust (next section).

In Fig.~\ref{fig:CCD}, we extend the colour-colour diagrams to optical
wavelengths. The datapoints show the photometry from only the
galaxy region, splitting the sample with respect to the projected
radial distance between the core -- where the starburst is taking
place (black dots, R$<$500\,pc) -- and the rest of the galaxy (grey
dots). We overlay in each panel three tracks showing simple stellar
populations (SSPs) at solar metallicity, with ages ranging from
30\,Myr (bottom-left of each panel) to 5\,Gyr. Crosses mark the ages
of 0.1, 0.5, 1 and 5\,Gyr, for reference. The blue solid lines follow
a dustless model, whereas the orange dotted (red dashed) lines
introduce a $0.5$\,mag ($0.75$\,mag) colour excess according to the
favoured \citet{Fitz:99} extinction law.

\begin{figure}
\includegraphics[width=8cm]{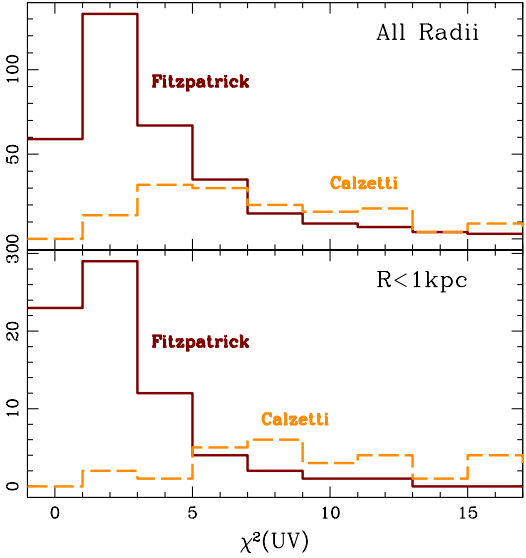}
\caption{
Comparison of the best fits between a Calzetti (dashed lines) and a
Fitzpatrick (solid lines) extinction law. Both optical and NUV
photometry is used to constrain the model parameters (as in
Fig.~\ref{fig:ages}), but in this case we use the NUV photometry to
define a new statistic, $\chi^2(UV)$. The results are restricted to
models with an acceptable {\sl total} reduced $\chi_r^2<5$ for either
the Calzetti, or the Fitzpatrick model), and with a corresponding
colour excess of E$_{\rm B-V}\simgt 0.1$\,mag.  The distributions for
the inner regions (R$<$1\,kpc) are shown in the bottom panel.  A wide
range of star formation histories are explored, including simple
stellar populations; exponentially decaying models and a two-burst
superposition (see text for details).  }
\label{fig:chi2FC}
\end{figure}

\subsection{Modelling the stellar populations}

We present here an extended analysis of the stellar populations,
exploring a wide range of star formation histories to derive a more
quantitative assessment of the extinction law. We run three grids of
models corresponding to different star formation histories: single
burst models (SSP); two-burst models (2SSP) and exponentially decaying
models (EXP).  Tab.~\ref{tab:params} shows the model parameters and
the sampling used in the grids. We use the population synthesis models
of \cite{BC03} to build the grids. In addition to the parameters
controlling the age distribution and the metallicity, we include an
additional parameter that describes the amount of reddening, via a
colour excess, E$_{\rm B-V}$, following either the extinction law
of \citet{Fitz:99} or \citet{Calz:01}.  For each aperture, we define a
$\chi^2$ in the usual manner, by comparing model (MOD) and measured
(OBS) data, using the SDSS $g$ band measurement as
normalization. Specifically, for each aperture, we define:

\begin{equation}
\chi^2\equiv \sum_{i=1}^8 \frac{(c_i^{\rm MOD}-c_i^{\rm OBS})^2}
{\sigma^2(c_i^{\rm OBS})},
\label{eq:chi2}
\end{equation}
where the $\{c_i\}$ represent the aperture colours, defined as 
\[
c_i = 
\left\{
\begin{aligned}
&g - X_i, \qquad i<8\\
&{\rm FUV}-{\rm UVW2}_c, \qquad i=8
\end{aligned}
\right.
\]
with $X_i=$\{UVW2,UVM2,UVW1,$u$,$r$,$i$,$z$\} ($i<8$), and
$\sigma(c_i^{\rm OBS})$ is the uncertainty of the $i^{\rm th}$ observed colour.
The last term ($i=8$) corresponds to the colour between the GALEX
FUV band and UVW2, where the comparison requires the UVW2 image to be convolved
to the (lower) resolution of the FUV passband, thus the notation
UVW2$_c$. Fig.~\ref{fig:ages} ({\sl left}) shows the
probability-weighted age ({\sl top}) and colour excess ({\sl bottom}),
for the \citet{Fitz:99} extinction law, corresponding to the single
burst (SSP) models. For reference, the age and dust reddening
estimates from \citet{RodMer:11} are included as grey shaded regions,
showing good agreement within the uncertainties. We note that
metallicity is treated in this paper as a nuisance parameter: although
metallicity cannot be constrained with this type of photometric data
alone, we need to explore a range of values of metallicity in order to
extract robust estimates of the age and dust content.  The
distribution of reduced $\chi_r^2$ values has a median of 0.70, and
$\sim$90\% of the data points have $\chi_r^2<3.0$ Hence, the reduced
values of $\chi^2$ stay around $\chi_r^2\sim$1--3, showing that the
fits are quite acceptable. Fig.~\ref{fig:SEDfit} illustrates the goodness
of fit with three typical cases at different galactocentric radii (as
labelled).  The error bars indicate the aperture photometry (the
horizontal error bars span the FWHM of the filter). The blue line
corresponds to the best fit spectrum in each case. The same apertures
are shown for a Fitzpatrick ({\sl left}) or a Calzetti ({\sl right})
extinction law.  Notice the significant mismatch of the NUV
photometry when using the Calzetti function, revealing the
presence of the NUV bump.

\begin{figure}
\includegraphics[width=9cm]{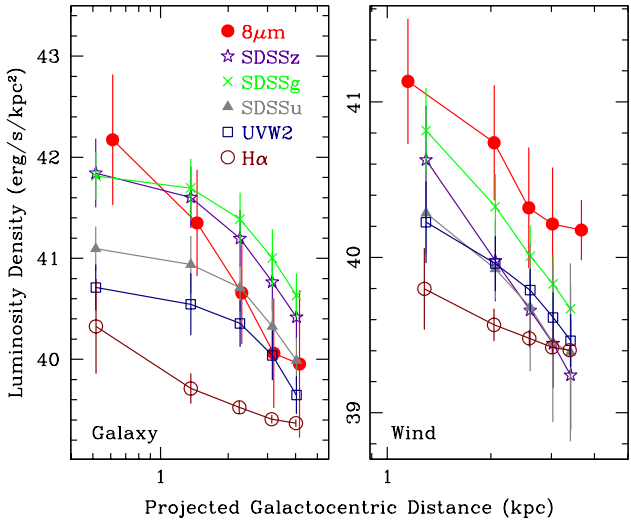}
\caption{Radial profile of luminosity density in the galaxy ({\sl left})
and wind region ({\sl right}) of M82. The points correspond to the median
value within bins in projected radial position (the binning is done at fixed
number of data points per bin). The RMS scatter in each bin is shown as
an error bar.}
\label{fig:RadProf}
\end{figure}

Fig.~\ref{fig:ages} ({\sl right}) shows the parameters of the best
fits for the composite population models: the top panels give the
fraction in young mass stars (f$_{\rm Y}$) with respect to the age of
the young component (t$_{\rm Y}$); whereas the bottom panels show the
exponential timescale ($\tau$), and the average stellar age for the
EXP models. In both cases, we split the sample into two panels ({\sl
left/right}), with respect to the projected galactocentric distance,
as labelled.  Note that in the outer region, the populations are well
described by a relatively homogeneous population with ages in the
range $0.5-1.5$\, Gyr, regardless of the model used. Within the central
kpc, the populations are significantly younger, although the modelling
is unable to distinguish whether the younger ages are caused by an
extended period of star formation (i.e. EXP models) or after a recent
burst (i.e. 2SSP models). For the general sample, the distribution
$\Delta\chi^2\equiv \chi^2_{\rm min,2SSP} - \chi^2_{\rm min,EXP}$ has
a mean of $-1.3$ with an RMS of $2.9$, i.e. slightly favouring a
two-burst scenario.

\begin{table*}
\caption{Model parameters used for the grid of star formation histories.
Each model is defined by the star formation rate, $\psi(t)$. In the
EXP models, we define $t_U$ as the age of the Universe at z=0.}
\label{tab:params}
\begin{tabular}{lccr}
\hline
\multicolumn{4}{c}{Single Burst (SSP): $\psi(t)\propto\delta(t-t_0)$}\\
\hline
Observable & Parameter & Range & Steps\\
\hline
Age & log(t$_0$/Gyr) & $-2\cdots +0.7$ & 64\\
Metallicity & $\log Z/Z_\odot$ & $-2\cdots +0.3$ & 16\\
Dust & E$_{\rm B-V}$ & $0\cdots 1$ & 32\\
\hline
\multicolumn{3}{r}{Number of models} & $32,768$\\
\hline
\multicolumn{4}{c}{Two Bursts (2SSP): $\psi(t)\propto\Big[
f_{\rm Y}\delta(t-t_{\rm Y}) + (1-f_{\rm Y})\delta(t-t_{\rm O})\Big]$}\\
\hline
Observable & Parameter & Range & Steps\\
\hline
Age, Old &(t$_{\rm O}$/Gyr) & $0.5\cdots 12$ & 16\\
Age, Young & log(t$_{\rm Y}$/Gyr) & $-2\cdots -0.3$ & 16\\
Young Mass Fraction & f$_{\rm Y}$ & $0\cdots 1$ & 16\\
Metallicity & $\log Z/Z_\odot$ & $-2\cdots +0.3$ & 16\\
Dust & E$_{\rm B-V}$ & $0\cdots 1$ & 32\\
\hline
\multicolumn{3}{r}{Number of models} & $2,097,152$\\
\hline
\multicolumn{4}{c}{Exponentially decaying rate (EXP): 
$\psi(t)\propto\exp\Big[-(t-t_{\rm FOR})/\tau\Big]$}\\
\hline
Observable & Parameter & Range & Steps\\
\hline
Formation Time & $\log t_S\equiv\log[(t_U-t_{\rm FOR})/{\rm Gyr}]$ & $-1\cdots +1$ & 32\\
Timescale & $\log(\tau/{\rm Gyr})$ & $-1\cdots +1$ & 32\\
Metallicity & $\log Z/Z_\odot$ & $-2\cdots +0.3$ & 16\\
Dust & E$_{\rm B-V}$ & $0\cdots 1$ & 32\\
\hline
\multicolumn{3}{r}{Number of models} & $524,288$\\
\hline
\multicolumn{3}{r}{TOTAL} & $2.65\times 10^6$\\
\end{tabular}
\end{table*}


Fig.~\ref{fig:chi2FC} provides a more quantitative confirmation of the
preference of a Milky Way (hereafter MW) type of extinction law with
respect to a ``bump-less'' \citet{Calz:01} extinction curve. It
compares the $\chi^2$ of the respective models, restricting the
results to NUV photometry (i.e. only using the information from FUV,
UVW2, UVM2 and UVW1 to define the statistic). For a meaningful
comparison, we consider only models that are acceptable, rejecting
those results with a total reduced $\chi_r^2>5$ in {\sl either} the
Calzetti or the Fitzpatrick cases. Only models with a best fit
reddening E$_{\rm B-V}>0.1$\,mag are included, since regions with low
reddening will not be sufficiently informative for the discrimination
between extinction laws. In order to determine whether there are
differences between the dust extinction law in the central starburst
or the periphery -- where a more standard MW-type behaviour is expected --
the figure is split between the central kpc (in projection) and the
outer regions. No significant changes are found, with an overall
preference towards a Fitzpatrick extinction. One could expect
that a projected measurement of an edge-on galaxy such as M82 would
introduce a MW-type extinction from the outer regions seen along the
line of sight. However, the fact that the ages and colour excess at R$<$1\,kpc
are clearly different from the rest would imply that the central burst
makes a large contribution to the photometry in many of the R$<$1\,kpc
apertures. Therefore, we can tentatively conclude that a Calzetti law
is not favoured in the starburst region of M82.

\begin{figure*}
\raisebox{-0.5\height}{\includegraphics[width=10.5cm]{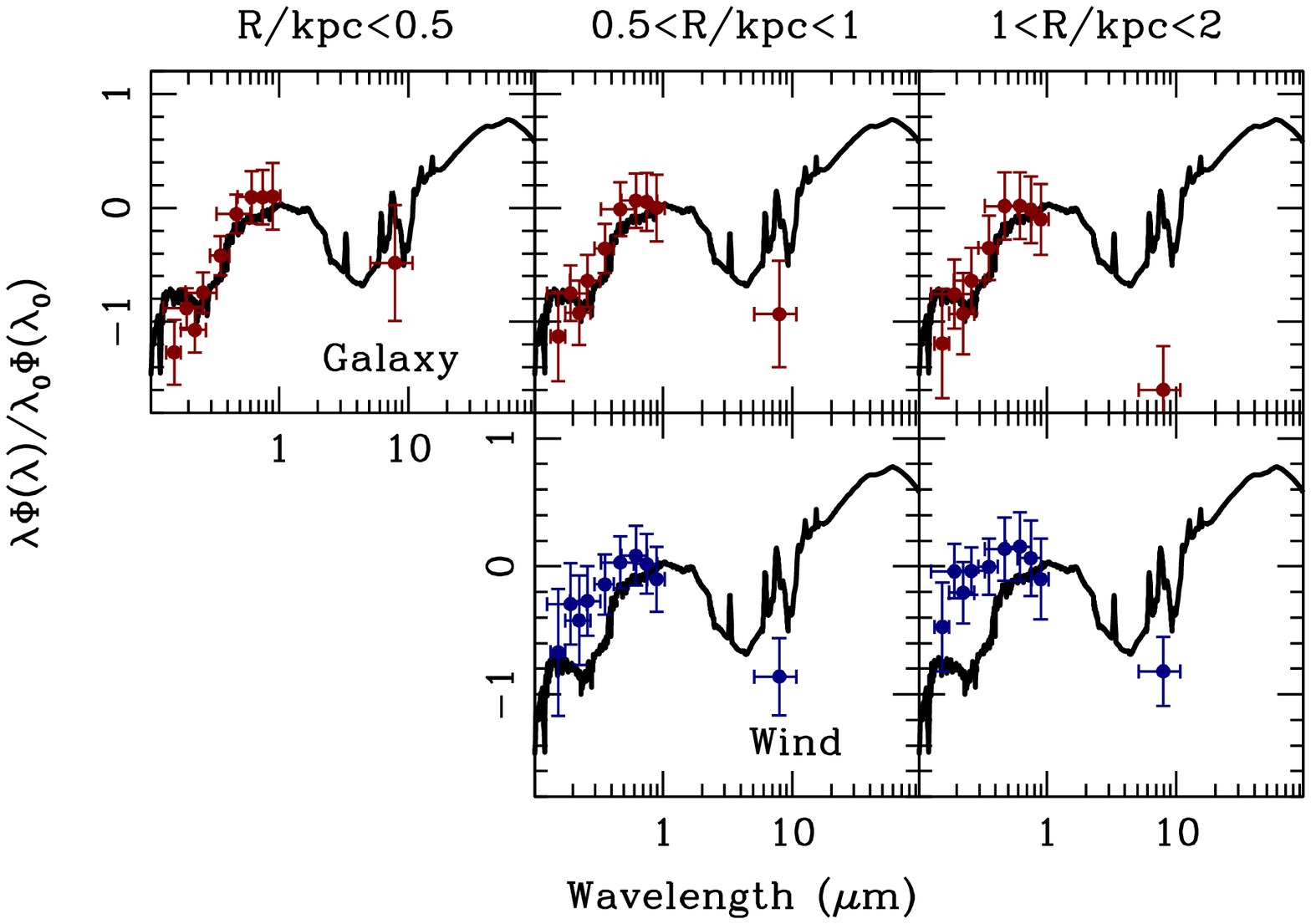}}
\raisebox{-0.5\height}{\includegraphics[width=7cm]{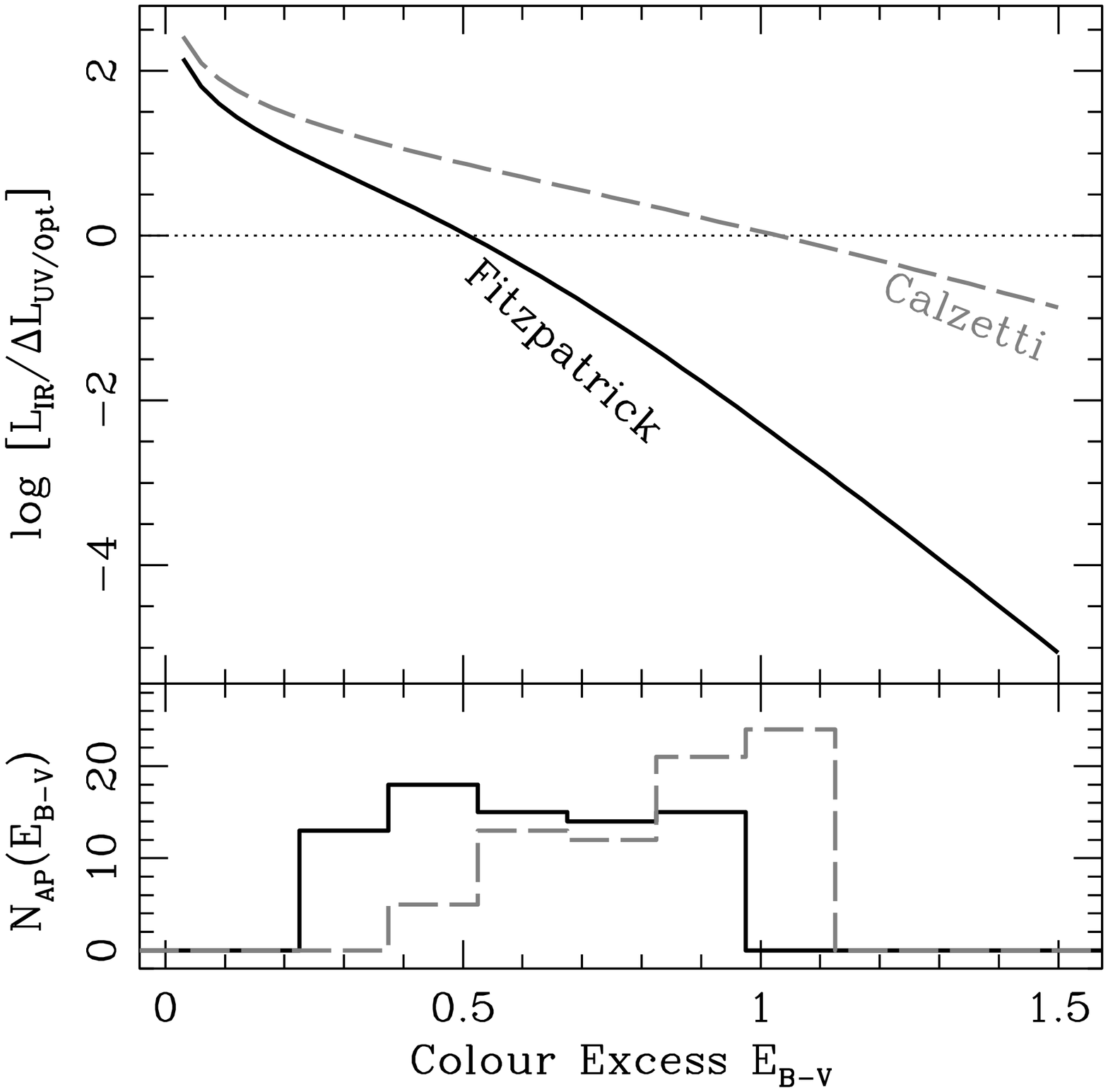}}
\caption{{\sl Left}: Comparison of the UV to IR spectral energy distribution 
in the galaxy (red points; top) and wind (blue points; bottom)
regions, with respect to projected radial distance to the centre. 
The data include GALEX/FUV; UVOT/UVW2, UVM2, UVW1; SDSS/u, g, r, i, z
and IRAC 8$\mu$m.  The horizontal error bars indicate the spectral
coverage of each passband, while the vertical error bars show the
RMS scatter within each radial bin.  The black line is a template of
the M82 spectrum \citep{Lonsdale:04}. All data are normalized to the
flux at $\lambda_0=0.8\mu$m. Notice the sharp decrease at 8$\mu$m in
the galaxy with increasing distance. In the wind region, the bluer SED
is caused by starlight scattered from dust entrained in the wind,
and the milder decrease in 8$\mu$m reveals intrinsic emission far
from the galactic plane.  {\sl Right}: (top) The ratio between the observed
IR luminosity ($\lambda>8\,\mu$m) and UV/Optical excess luminosity (estimated
between the observed and the unreddened spectra in
$0.1<\lambda/\mu$m$<1$) is shown with respect to colour excess
for the Fitzpatrick (black solid line) or Calzetti (grey dashed line)
attenuation laws. (Bottom) The histograms represent the best-fit values from 
the SSP models for the R$<$1\,kpc apertures (see Fig.~\ref{fig:ages}).}
\label{fig:SED}
\end{figure*}

Fig.~\ref{fig:RadProf} compares the radial profile in the galaxy and
wind regions at different wavelengths, from UVW2 to 8$\mu$m, including
H$\alpha$.  There is a very sharp decrease of the 8$\mu$m data in the
galaxy region, in contrast with a milder gradient in the wind region,
reflecting an additional contribution at 8$\mu$m from intrinsic
emission of the dust entrained in the wind material. In addition, the
gradient at shorter wavelengths is shallower in the wind region, a
consequence of the wavelength-dependent scattering of light from the
dust component. This effect is easier to visualize in
Fig.~\ref{fig:SED} ({\sl left}), where we show the average value of
the flux in several regions in the galaxy ({\sl top}) and the wind
({\sl bottom}). For reference, we include the integrated spectral
energy distribution of M82 from the templates
of \citet{Lonsdale:04}. Notice the good agreement in the central part
of the galaxy, where the starburst phase is taking place. As we move
further out along the galaxy disc, the PAH-dominated emission at
8$\mu$m drops sharply. In contrast, the wind region shows a smaller
decrease of the 8$\mu$m flux with radial distance, and the NUV-optical
spectrum becomes significantly bluer.  In the next section, we will
relate this trend with the properties of the dust entrained in the
wind.

The comparison of the NUV/optical and FIR spectrum of M82 in
Fig.~\ref{fig:SED} ({\sl left}) raises the issue of the energetic
balance between the light ``removed'' by dust from the NUV/optical
region, and the FIR emission -- which corresponds to energy
re-radiated by the heated dust. In other words, could M82 hide a
heavily dust-enshrouded starburst whose FIR emission is unaccounted
for in the NUV/optical window? Fig.~\ref{fig:SED} ({\sl right}) shows the
result for a simple model, where the original template of M82, shown in
the leftmost panels, is dereddened according to the colour excess,
E$_{\rm B-V}$, which is taken as a free parameter (horizontal
axis). The vertical axis represents the ratio between the observed
total IR luminosity (measured at $\lambda>8\,\mu$m) and the {\sl
excess} luminosity in the UV and optical spectral region. The latter
is obtained by subtracting the observed energy in the UV-optical range
($0.1<\lambda/\mu{\rm m}<1$) from the one corresponding to the
dereddened spectrum for a given E$_{\rm B-V}$, using the standard
Fitzpatrick (solid line) or Calzetti (dashed line) extinction
law. Note that the L$_{\rm IR}$/$\Delta$L$_{\rm UV/Opt}=1$ case (horizontal
dotted line) corresponds to the 1:1 balance between the UV/Optical light
absorbed by dust and the dust emission at longer wavelengths.  The
histograms ({\sl bottom}) are the distributions of best-fit 
$E_{\rm B-V}$ obtained in the modelling of the stellar populations of
the R$<$1\,kpc apertures (see, e.g. Fig.~\ref{fig:ages}), confirming
that the reddening thus obtained is consistent. Therefore, within
uncertainties, all dust-enshrouded light is accounted for.

\section{Dust in the superwind}
\label{Sec:DustModel}

\subsection{Probing the dust properties}

\begin{figure}
\includegraphics[width=8cm]{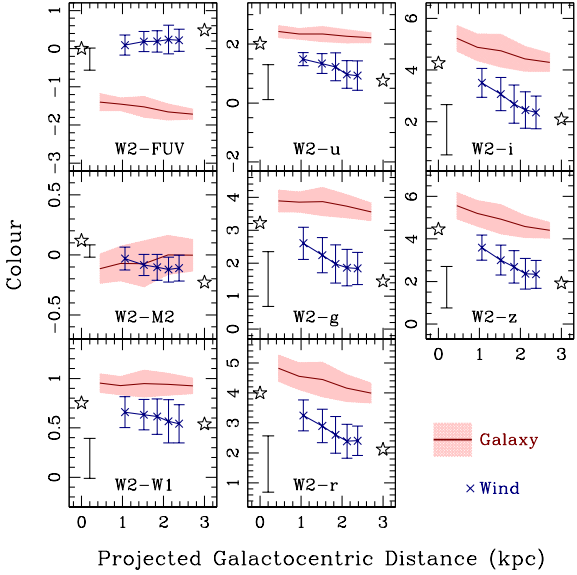}
\caption{Radial plots of the measured colours in the wind (blue crosses) 
and galaxy areas (red line and shaded regions). The colours,
UVW2$-X$, with $X$ ranging from UVM2 to SDSS-$z$, are given as the
the median values, binned at a fixed number of data points per bin.
The error bars and the extent of the shaded regions give the RMS
spread within each bin. The stars show the extrapolated colours
in the wind region at R=0 and 3\,kpc. The error bar on the left-hand side
of each panel is the predicted best-fit SSP models at R$<$1\,kpc
when the dust screen is removed (see text for details).}
\label{fig:WindProf}
\end{figure}

Fig.~\ref{fig:WindProf} shows the UVW2$-X$ radial colour profiles
in the wind region (blue crosses along with the RMS scatter as error
bars), with $X$ ranging from GALEX/FUV to the SDSS $z$ band, as
labelled.  For reference, the flatter radial profiles of the colours
in the galaxy region are shown in each panel as a red line with a
shaded area extending over the observed RMS scatter. The colour
gradients in the galaxy increase slightly with wavelength, as expected
from typical variations in stellar age, metallicity and dust
reddening. However, the gradients in the wind region are significantly
steeper. Extrapolations of the colour in the wind region at
galactocentric distances of 0 and 3\,kpc are shown as star symbols.
Light from the wind region can be explained only by scattering, shocks
or photoionisation. However, the recent analysis of \citet{Hoopes:05},
using GALEX UV and H$\alpha$ photometry reject the last two, leaving
dust scattering as the main cause for the observed light.

In this section, we explore the constraints one can enforce on the
distribution of dust via the wavelength dependent change between the
light in the central part of the galaxy -- where the emission
originates -- and the outer region of the wind -- where the
contribution is almost exclusively caused by dust scattering. We
determine the change induced by scattering from the colour change
between the measurements at projected galactocentric distances of R=0
and 3\,kpc. As a reference, the error bars on the left-hand side of
each panel in Fig.~\ref{fig:WindProf} give the predicted colours of a
model for all measurements at R$<$1\,kpc, for which the dust screen
from the best-fit result is removed, keeping the age and metallicity
unchanged. The large scatter is a result of the wide range of values
of E$_{\rm B-V}$ in the central region of the galaxy (see
Fig.~\ref{fig:ages}, {\sl left}). These colours are in most cases
bluer than the photometric measurements in the wind, showing that the
illumination source from the starburst must be significantly affected
by dust.  The variation of the colours between R=0 and R=3\,kpc in the
wind region are shown with respect to wavelength in
Fig.~\ref{fig:DustScat} (filled dots), where the error bars correspond
to the RMS scatter. For reference, a generic dust scattering law is
assumed:
\begin{equation}
\sigma_{\rm scat} \propto \lambda^{-x}
\label{eq:scat}
\end{equation}
The three lines in the figure represent the expectation for
three choices of $x$, as labelled.
Assuming the incident light originates in the central
starburst, we use simple stellar population models with the same age and
reddening properties as in the central regions of the galaxy (see
Fig.~\ref{fig:ages}), and modify the spectrum according to
equation~\ref{eq:scat}.  The inset gives the probability
distribution function keeping $x$ as a free parameter, where we find
$x = 1.53\pm 0.17$ ($1\sigma$ error bar).
We note that in the limit of small particle size
(Rayleigh scattering), $x=4$, whereas in the opposite regime
of dust grains much larger than the incident wavelength, $x=0$ is expected
\citep[see, e.g.,][]{Draine:book}.

\subsection{A simple dust scattering model}
The radiation from the galactic wind cone of M82 is probably scattered
light from the starburst in the central region by
free electrons, molecules and dust entrenched in the wind outflows.
Scattering by free electrons is practically Thomson scattering, whose
cross-section is independent of the wavelength of the incident
radiation.  Scattering by molecules can be described as a Rayleigh
scattering process -- given the smaller size of the molecules with
respect to the wavelength of the optical/NUV radiation.  The cross
section of Rayleigh scattering scales as $\lambda^{-4}$, thus the
process preferably scatters the higher-frequency radiation away from
the incident rays.  Scattering by dust particles is more complicated.
However, as an approximation, if we neglect the thermodynamics of the
process, we may employ the Mie prescription, in which the scattering
dust kernels are modelled by dielectric or metallic spheres and the
incident radiation as waves. There are two distinctive regimes in this
scattering process, depending on the ratio between the wavelength of
the incident radiation and the size of the scattering spheres.  The
critical wavelength ($\lambda_c$) that divides the two regimes is
therefore determined by the characteristic size of the scattering 
sphere ($a_c$).
Radiation with $\lambda>\lambda_c$, will be scattered with the
Rayleigh scaling.  For $\lambda<\lambda_c$ -- if ignoring resonant
features, the cross-section is practically independent of the
wavelength of the incident radiation, and the process will be
described by Thomson scattering.  Note that these different regimes
have distinguishable angular dependence.  However, because of the
specific geometrical setting of the source and the
orientation of the wind cone in M82,  the
angular dependence of the scattering processes does not play a very
significant role in determining the wavelength dependence of the
observed radiation in the weak scattering limit.

We now show that the observed $\propto\lambda^{-1.53}$ dependence of
the radiation in the NUV/optical data can be explained by a simple
dust scattering model. We assume an incident radiation originating
from a point source in the core of the galaxy.  The observed light in
the wind region results from the scattering of this radiation by the dust
particles in the wind cone. We consider spherical dust particles with
some distribution of sizes. Even though the particle density can vary
with location, we enforce the same distribution of dust grain size
everywhere in the wind cone.  As the sizes of the dust particles
relative to the wavelengths of the incident radiation determine
whether the scattering is in the Rayleigh regime or in the Thomson regime,
the scattering optical depth is wavelength dependent. One can define  a critical
wavelength $\lambda_c$ that reflects the transition from one regime to
the other.  If all dust grains had the same size, we would expect a
trivial spectral break in the wavelength dependence of the scattered
light at $\lambda_c$ -- see panel a) of Fig.~\ref{fig:model}, where
$\lambda_c=2000$\,\AA\ is assumed. An ensemble of dust particles with
different sizes smear the spectral break into a broad region,
depending on the particle size distribution.

In the weak scattering approximation, the intensity scattered into the
line-of-sight $I_{\rm sc}$ from the incident rays, with intensity
$I_{0}$, may be expressed as:

\begin{eqnarray}
 I_{\rm sc}(\lambda) & \approx &  \tau_{\rm sc}(\lambda)\, \left(
    I_{0}(\lambda)\, e^{-\tau_{\rm ex}(\lambda)} \right)\bigg\vert_{\rm {\mathbf r}} \ , 
\end{eqnarray}

\noindent
where $\tau_{\rm sc}$ is the scattering optical depth and $\tau_{\rm
ex}$ is the extinction depth evaluated at the projected location
${\mathbf r}$.  This expression takes into account the attenuation of
the incident radiation due to scattering and absorption when
propagating through the scattering wind cone.  In the spectral
calculation, the optical depths $\tau_{\rm sc}(\lambda)$ and
$\tau_{\rm ex}(\lambda)$ can be derived using a parametric
prescription with the weights of the critical wavelength $\lambda_{c}$
determined by the convolution of the grain size distribution. We can
therefore compute the spectrum of the scattered radiation using a
reference incident spectrum, such as the SED of the star-forming
region in the core of M82 (red circles in Fig.~\ref{fig:SED}, left
panel, at R$<$1\,kpc).

\begin{figure}
\includegraphics[width=8cm]{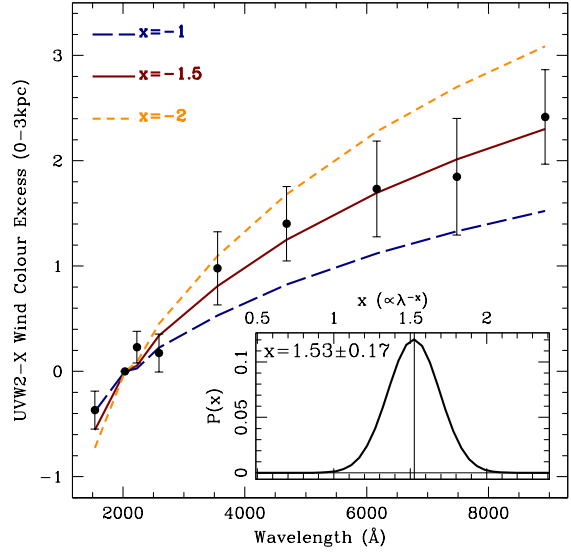}
\caption{Colour variation between the galaxy and the wind against 
wavelength. The black filled circles are the observations, including
the RMS scatter as error bars. The three lines correspond to 
different dust scattering laws, proportional to $\lambda^{-x}$. The
inset shows the probability distribution function for the power law
index $x$.}
\label{fig:DustScat}
\end{figure}

In panel b) of Fig.~\ref{fig:model}, we show that the observed
$\propto\lambda^{-1.53}$ dependence (Fig.~\ref{fig:DustScat}) can be
obtained by a dust model for the wind cone with a size distribution
$n(a) \propto a^{-2.5}$, where $a$ is the dust grain radius. In
addition, the particles must have upper and lower size limits such
that the critical wavelengths corresponding to these size limits fall
outside of the wavelength range covered (i.e. limited by the grey
shaded regions in Fig.~\ref{fig:model}). An almost perfect power-law
is obtained, with similar wavelength dependence to our
results. Alternatively, the grain size distribution can be flatter,
i.e. $n(a) \propto a^{-\gamma}$, where the power law index $\gamma <
2.5$, with an additional constraint on the maximum size of the dust
particles.  The critical wavelengths corresponding to the largest
grains must lie within the wavelength range covered by the
observations.  In this case, we would not obtain a power law but the
result would be compatible with the observed $\propto\lambda^{-1.53}$
behaviour. This dependence is due to the presence of a broad
transition region between the Rayleigh scattering regime (with
$\lambda^{-4}$ dependence) and the Thomson scattering regime (practically
independent of $\lambda$) within the spectral range covered.  Panel c)
of Fig.~\ref{fig:model} demonstrates that an acceptable fit can be
generated for a dust size distribution $n(a) \propto a^{-1.8}$ with
the additional constraint of an upper size limit ($a_{\rm MAX}$), that
gives a critical wavelength $\lambda_{c}(a_{\rm
MAX})\simeq\,1\mu$m.

\begin{figure}
\includegraphics[height=8cm]{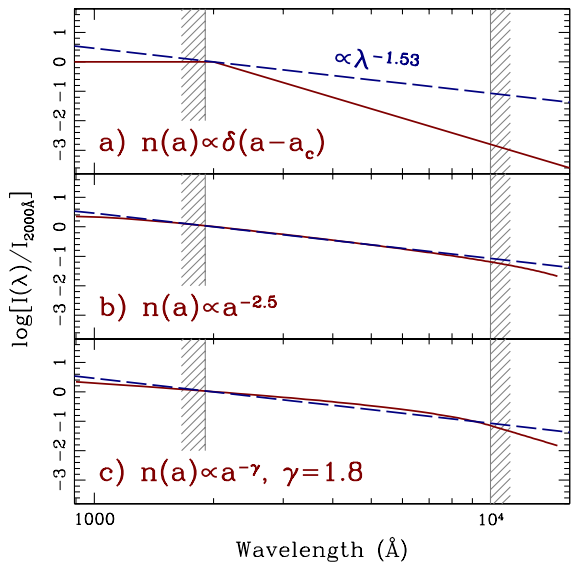}
\caption{Predictions for the intensity of light
scattered by the three simple dust models considered
in \S\ref{Sec:DustModel} (see text for details). The grey shaded
regions mark the ends of the observed wavelength coverage, and the
blue dashed line in all three panels corresponds to a fiducial
$\propto\lambda^{-1.53}$ behaviour, from the observational constraints
(see Fig.~\ref{fig:DustScat}).  }
\label{fig:model}
\end{figure}

\section{Conclusions}
\label{Sec:Conc}

M82 is the nearest starburst galaxy, allowing us to explore with a
high level of detail the various processes of this important phase of
galaxy evolution. We present here a study of deep NUV images taken by
the UV/Optical Telescope on board the {\sl Swift} observatory. We
combine them with additional UV, optical and IR archival data, to
explore the properties of the stellar populations in the galaxy, and
the dust entrained in the supernov\ae-driven wind. The NUV
colour-colour diagram -- especially sensitive to the presence of the
2175\,\AA\ bump -- reveals a strong rejection of traditional
extinction curves used for starburst galaxies \citep{Calz:01} which
lack a bump. The standard Milky Way extinction \citep[e.g.][]{Fitz:99}
is favoured (Fig.~\ref{fig:NUVccd}). The stellar populations reveal a
very young core, with luminosity-weighted ages $\simlt $100\,Myr and
large extinction (E$_{\rm B-V}\simgt $0.5\,mag) at projected
galactocentric distances R$<$1\,kpc. In the outer regions, the galaxy
has an overall homogeneous distribution of 0.7--1\,Gyr old
populations, with lower, but significant (E$_{\rm B-V}\sim$0.2\,mag)
colour excess (Fig.~\ref{fig:ages}). In the wind region, the spectral
energy distribution is bluer, and the PAH-dominated emission at
8$\mu$m does not decrease with galactocentric distance as sharply as
along the disc (Fig.~\ref{fig:RadProf}). These two trends reflect the
contribution from dust scattering, and from intrinsic dust emission,
respectively. In addition, the energy balance between the
observed IR ($\lambda>8\mu$m) emission and the UV/optical
($\lambda<1\mu$m) energy absorbed by dust according to our best-fit
models, suggests that there is no excess energy  in the
form of a heavily dust-enshrouded starburst (see Fig.~\ref{fig:SED},
{\sl right}).

By comparing the colours in the wind region over a wide separation in
galactocentric distance, $\Delta R=3$\,kpc, we quantify the 
wavelength-dependent scattering. We obtain a behaviour $\propto\lambda^{-1.5}$
(Fig.~\ref{fig:DustScat}), implying either a distribution of dust
grain sizes as $n(a)\propto a^{-2.5}$, or a flatter distribution,
e.g. $n(a)\propto a^{-1.8}$, along with an upper size limit that
results in a critical wavelength within the spectral coverage of the
instrumentation. Although detailed values of this size limit would
require information about the grain composition, this result would
suggest that only small grains are entrained in the supernov\ae-driven
wind.



\label{lastpage}

\end{document}